International
Virtual
Observatory
Alliance

# Sky Event Reporting Metadata (*VOEvent*) Version 2.0

## IVOA Recommendation 27 June 2011

**Editors:**

Rob Seaman, seaman@noao.edu
Roy Williams, roy@caltech.edu

**Authors:**

Rob **Seaman**, National Optical Astronomy Observatory, USA
Roy **Williams**, California Institute of Technology, USA
Alasdair **Allan**, University of Exeter, UK
Scott **Barthelmy**, NASA Goddard Spaceflight Center, USA
Joshua S. **Bloom**, University of California, Berkeley, USA
John M. **Brewer**, Yale University, USA
Robert B. **Denny**, DC-3 Dreams SP, USA
Mike **Fitzpatrick**, National Optical Astronomy Observatory, USA
Matthew **Graham**, California Institute of Technology, USA
Norman **Gray**, University of Glasgow, UK
Frederic **Hessman**, University of Gottingen, Germany
Szabolcs **Marka**, Columbia University, USA
Arnold **Rots**, Harvard-Smithsonian Center for Astrophysics, USA
Tom **Vestrand**, Los Alamos National Laboratory, USA
Przemyslaw **Wozniak**, Los Alamos National Laboratory, USA

## Abstract

VOEvent [20] defines the content and meaning of a standard information packet for representing, transmitting, publishing and archiving information about a transient celestial event, with the implication that timely follow-up is of interest. The objective is to motivate the observation of targets-of-opportunity, to drive robotic telescopes, to trigger archive searches, and to alert the community. VOEvent is focused on the reporting of photon events, but events mediated by disparate phenomena such as neutrinos, gravitational waves, and solar or atmospheric particle bursts may also be reported.

Structured data is used, rather than natural language, so that automated systems can effectively interpret VOEvent packets. Each packet may contain zero or more of the "who, what, where, when & how" of a detected event, but in addition, may contain a hypothesis (a "why") regarding the nature of the underlying physical cause of the event.

Citations to previous VOEvents may be used to place each event in its correct context. Proper curation is encouraged throughout each event's life cycle from discovery through successive follow-ups.

VOEvent packets gain persistent identifiers and are typically stored in databases reached via registries. VOEvent packets may therefore reference other packets in various ways. Packets are encouraged to be small and to be processed quickly. This standard does not define a transport layer or the design of clients, repositories, publishers or brokers; it does not cover policy issues such as who can publish, who can build a registry of events, who can subscribe to a particular registry, nor the intellectual property issues.

## Table of Contents

# Status of this Document

This is the Proposed Recommendation of VOEvent v2.0. The first release of the near final document was 2011 March 07.

This document describes an update to the IVOA VOEvent recommendation for the representation of celestial transient event metadata. It has been developed by the VOEvent Working Group. Changes from v1.1 are described below (in section 1.1).

This is an IVOA Proposed Recommendation made available for public review. It is appropriate to reference this document only as a recommended standard that is under review and which may be changed before it is accepted as a full recommendation.

The latest version (currently v1.11) of the internationally adopted VOEvent Recommendation (standard) is available from http://ivoa.net/Documents/latest/VOEvent.html. The list of current IVOA Recommendations and other technical documents can be found at http://www.ivoa.net/Documents/.

VOEvent is an IVOA standard, which means that it fits into a rich matrix of other IVOA standards: the image below shows where VOEvent fits into the broader IVOA architecture. VOEvents inherit much of the structure and semantics of **VOTable**, including the **UCD** scheme for semantics of quantities. VOEvent takes space-time coordinates from the **STC**, and it uses the URI semantics of the IVOA **Vocabulary** effort. IVOA **Identifiers** are used for events and their parent streams and servers, and both these latter will be described by IVOA **Resource Metadata** and stored in the registry.

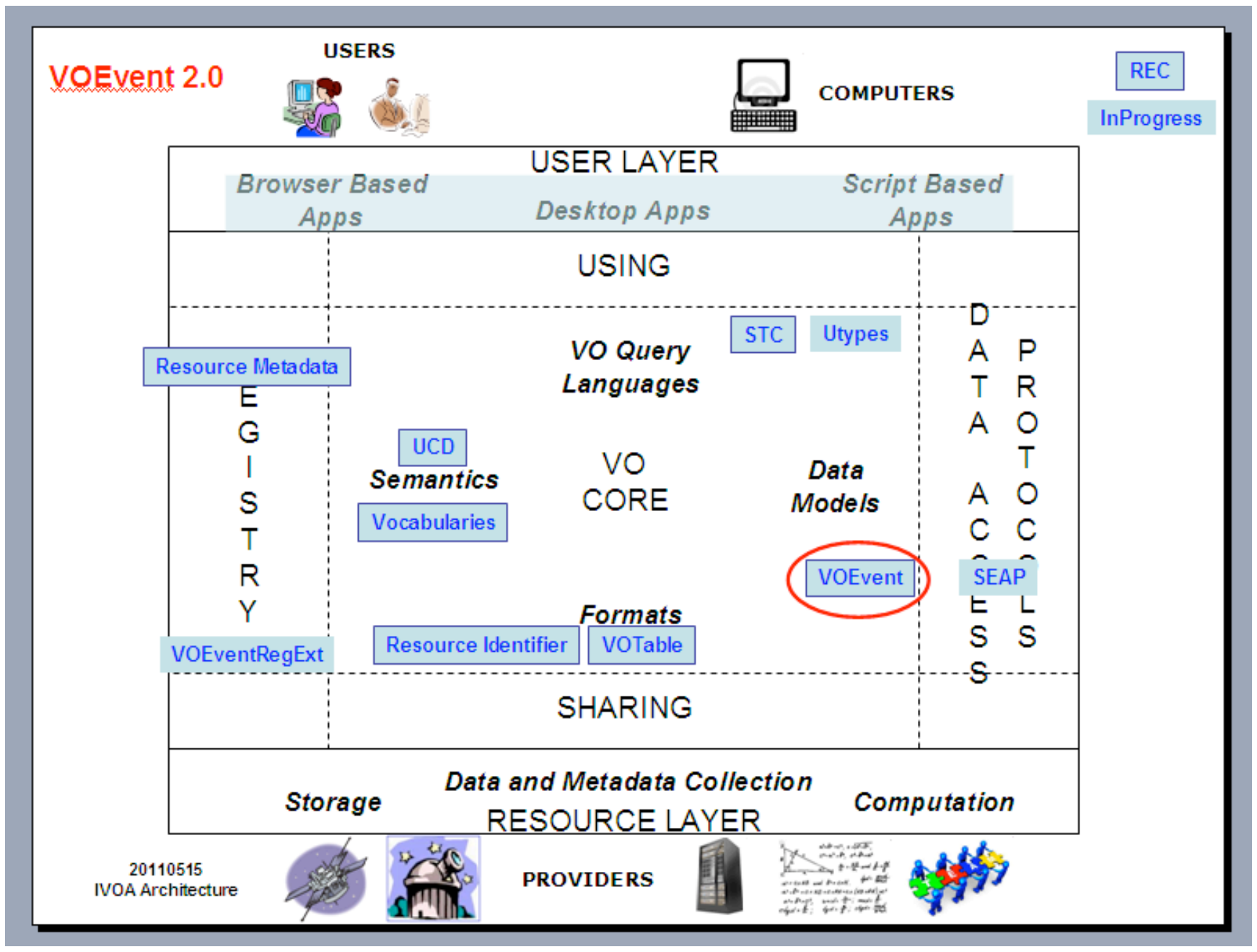

## 1. Introduction

Throughout human history, unexpected events in the sky have been interpreted as portents and revelations. Modern curiosity seeks to use such transient events to probe the fundamental nature of the universe. In decades to come the scientific study of such events will be greatly extended, with new survey telescopes making wide-area systematic searches for time-varying astronomical events, and with a large number of robotic facilities standing ready to respond. These events may reflect purely local solar system phenomena such as comets, solar flares, asteroids and Kuiper Belt Objects, or those more distant such as gravitational microlensing, supernovae and Gamma-Ray Bursts (GRBs). Most exciting of all may be new and unknown types of event, heralding new horizons for astrophysics. Searches for astrophysical events are taking place at all electromagnetic wavelengths from gamma-rays to radio, as well as quests for more exotic events conveyed by such means as neutrinos, gravitational waves or high-energy cosmic rays.

For many types of events, astrophysical knowledge is gained through fast, comprehensive follow-up observation — perhaps the immediate acquisition of the spectrum of a suspected optical counterpart, for example — and in general, by observations made with instruments in different wavelength regimes or at different times. To satisfy these needs, several projects are commissioning robotic telescopes to respond to digital alerts by pointing the telescope and triggering observations in near real-time and without human intervention. These include, for instance, Skyalert[5] in the USA, and RoboNet-1.0 [12] and eStar[3] in the United Kingdom. Automated systems may also query archives and initiate pipelines in response to such alerts.

Many projects have been conceived — some now in operation — that will discover such time-critical celestial events. These include a large number of robotic survey and monitoring telescopes with apertures from tens of centimeters to tens of meters, large-field survey projects like CRTS [8], PTF [31], Pan-STARRS [9] and LSST [7], satellites like Swift and Fermi [11], and more singular experiments like LIGO [6]. The community has demonstrated that robotic telescopes [13] can quickly follow-up events using the standard outlined in this document. In the past, human-centric event alert systems have been very successful, including the Central Bureau for Astronomical Telegrams (CBAT) [2] and the Astronomer's Telegram (ATEL) [1], but these systems use predominantly natural-language text to describe each event, and do not have sophisticated selection criteria for subscribers. The GRB Coordinates Network (GCN) [4] reports one of the most fruitful event streams of current times, and its events are transmitted very successfully for follow-up within seconds or minutes. With VOEvent, we leverage the success of GCN by making it interoperable with other producers of events, and by generalizing its transport mechanisms.

A much larger rate of events can be expected as new facilities are commissioned or more fully automated. These rates indicate events that must be handled by machines, not humans. Subscribing agents must be able to automatically filter a tractable number of events without missing any that may be key to achieving their goals. In general, the number of pending events from a large-scale survey telescope (such as LSST) that are above the horizon at a given observatory during a given observing session may be orders of magnitude larger than a human can sift through productively. Selection criteria will need to be quite precise to usefully throttle the incoming event stream(s) — say — "*give me all events in which a point source R-band magnitude increase of at least -2.0 was seen to occur in less than four hours, that are located within specified molecular column density contours of a prioritized list of galactic star forming regions*". In practice the result of complex queries such as these will be transmitted through intermediary "brokers" — which will subscribe to VOEvent-producing systems and provide filter services to client groups ("subscribers") via specialized VOEvents. Filtering will often be based on coincidence between multiple events. A gravitational wave detector may produce a large number of candidate events, but the interesting ones may be only those that register with multiple instruments.

Handling the anticipated event rates quickly and accurately will require alert packets to be issued in a structured data format, not natural language. Such a structured discovery alert — and any follow-up packets — will be referred to as a VOEvent. VOEvent will rely on XML schemata [29] to provide the appropriate structured syntax and semantics. These schemata may be specific to VOEvent or may implement external schemata such as the IVOA's Space-Time Coordinate (STC) metadata specification [17]. Some of the VOEvent structure is provided by this document, for example the meaning of the <Who> and <Date> elements; however other structure is provided by the author of the event stream, who might define, for example, what the <peak_energy> and <energy_variance> parameters mean when supplied with one of those events.

VOEvent is a pragmatic effort that crosses the boundary between the Virtual Observatory and the larger astronomical community. The results of astronomical observations using real telescopes will be expressed using the IVOA VOEvent standard and disseminated by registries and brokers within and outside the VO. Each event that survives rigorous filtering can then be passed to other telescopes to acquire follow-up observations that will confirm (or deny) the original hypothesis as to the classification of the object(s) or processes that generated that particular VOEvent in the first place. This must happen quickly (often within seconds of the original VOEvent) and must minimize unnecessary expenditures of either real or virtual resources.

VOEvent is *transport neutral*, but deploying and operating a robust general-purpose network of interoperating brokers has always been a high-priority issue. Various special-purpose networks and prototype networks for the global VOEventNet have been deployed and operated. See references under SkyAlert [5] and Transport [33] for two options.

Following the Abstract and Introduction, this document contains a discussion of appropriate VOEvent usage in §2. Section 3 is the heart of the document, conveying the semantics of a VOEvent packet. Explicit examples of VOEvent packets are in §4, and linked references in §5.

## 1.1 Changes from VOEvent 1.1

- The concept of event stream is introduced in section 2.2, this is new in VOEvent 2.0. The stream metadata acts as a template for the events in the stream, and is registered with the VO registry.
- The section on transport of VOEvents has been removed, so it can be handled in its own standards process.
- The section on Registry enhancements to support VOEvent has been expanded and clarified.
- The <Param> elements can now have <Description> and <Reference> elements
- The value of a <Param> element can now be expressed as an element in addition to an attribute.
- The <Param> element now has an attribute "dataType" to express the meaning of the parameter value (int, float, string).
- There is a new <Table> element to express simple tables, see section 3.3.3.
- The <Param> and <Field> elements may have an attribute "utype" to express how it fits into an IVOA data model.
- The VOEvent packet structure still conforms to the IVOA *Space-Time Coordinates* standard, but there is a new, simplified schema for these elements that is completely within the VOEvent schema.
- GPS time is now a valid time system for VOEvents
- The semantic implication of a <Citation> element is clarified: section 3.7
- The <Reference> element has a more sophisticated notion of *meaning*; it is a general URI reference to a wide range of possible content, rather than just a simple HTML link, and there is also a *mimetype* attribute.

# 2. Usage

This document defines the syntax and semantics of an alert packet known as VOEvent [20]. In this document, the word *packet* will refer to a single, syntactically complete, VOEvent alert or message, however transmitted or stored. The transmission of such a packet announces that an astronomical "event" has occurred, or provides information contingent on a previous VOEvent through a citation mechanism. The packet may include information regarding the "who, what, where, when & how" of the event, and may express "why" hypotheses regarding the physical cause of the observed event and the likelihood of each of these hypotheses.

## 2.1 Publishing VOEvent Packets

VOEvent packets express sky transient alerts. VOEvent users subscribe to the types of alerts pertinent to their science goals. The following roles define the interchange of VOEvent semantics:

- An ***Author*** is anyone (or any organization) creating scientific content suitable for representation as a sky transient alert. An author will typically register with the IVOA registry, so that the **<Who>** element of VOEvent packets can be small and reusable, expressing only the IVOA identifier needed to retrieve the contact information for the author. An authoring organization or individual may often rely on autonomous systems to actually create and transport the individual alert messages.

- A ***Publisher*** receives alerts from one-or-more authors, and assigns a unique identifier to each resulting packet. Either the author or the publisher generates the actual XML syntax of the event, but the publisher is responsible for the validity of the packet relative to the VOEvent schema. Publishers will register with the IVOA registry as described below.

- A ***Repository*** subscribes to (or is party to the original creation of) one or more VOEvent streams, persists packets either permanently or temporarily, and runs a service that allows clients to resolve identifiers and apply complex queries to its holdings. A given packet had one Publisher but may be held in more than one Repository. Public repositories will register with the IVOA registry.
- A ***Subscriber*** is any entity that receives VOEvent packets for whatever purpose. Subscribers can find out how to get certain types of events by consulting the lists of publishers and repositories in the IVOA registry. A subscription is a filter on the stream of events from a publisher: the subscriber is notified whenever certain criteria are met. For example, the filter may involve the curation part of the event (*e.g.*, "*all events published by the Swift spacecraft*"), its location ("*anything in M31*"), or it may reference the detailed metadata of the event itself ("*whenever the cosmic ray energy is greater than 3 TeV*").
- A ***Broker*** or ***Relay***, also sometimes known as a *Filter*, is any combination of the atomic roles of Publisher, Repository, or Subscriber that also offers arbitrary application-level functionality. See the IVOA VOEvent Transport Note [20] for further discussion.

## 2.2 VO Identifiers (IVORNs)

VOEvent benefits from the IVOA identifiers developed for the VO registry. Such an identifier is called an *IVORN*, that is, an *International Virtual Observatory Resource Name*. It is required to begin with "***ivo://***", and will stand in for a particular packet. A *registered* VOEvent packet is one that has a valid identifier — meaning that a mechanism exists that can resolve that identifier to the full VOEvent packet. VOEvent identifiers thus provide a citation mechanism — a way to express that one VOEvent packet is a *follow-up* in some fashion of a previous packet. For these reasons, VOEvent packets will often contain VO identifiers [15]. These take the general form ***ivo://authorityID/resourceKey#local_ID***, and are references to metadata packets that may be found at a VO registry or VOEvent repository. There are several types of metadata schema that the registry can hold. For the purposes of VOEvent, the principal schemata are:

- **VOEvent**: the metadata packet for an alert resulting from the observation of a transient celestial event. This schema is defined in this document.
- **VOEventStreamRegExt**: the metadata packet for a stream of VOEvents, including information about who is running it, the parameters that may be included and their meaning. The VOEventStreamRegExt may also be shortened to just 'stream', they mean the same thing.
- **VOEventServerRegExt**: the metadata packet to describe a service that provides VOEvents, which may be past events from a repository query service, or may be events sent in near real-time from a subscription service. The server definition includes the list of streams whose events are kept, the service endpoint to query the repository, or the endpoint to resolve a VOEvent identifier.
- **Author Organization**: these metadata [16] describe an author, including contact information and a description of the project. The VOEvent **<Who>** element contains either a reference to an author's IVORN or explicit contact information sufficient to describe the author.

When such an identifier is *resolved*, it means that the VOEvent metadata packet is obtained in exchange for the identifier. Such resolution happens through the global, distributed IVOA registry in stages. The registry is queried to locate a repository holding the relevant packet, and then the repository is queried for the packet itself. The part of the IVORN before the "#" symbol points to the *VOEventStreamRegExt* of which the event is a member; the whole IVORN (that includes the local_ID) points to the event itself. Thus VOEvent identifiers serve two purposes; they contain a stream identifier, then the "#" sign, then the local reference within that stream.

This is a key point in understanding VOEvent identifiers: **The Event identifier also expresses the Stream identifier.** For example:

- *ivo://nvo.caltech/voeventnet/catot#1004071150784109051*

  This identifier points to a specific VOEvent (number *1004071150784109051*) that is an instance of the stream called *ivo://nvo.caltech/voeventnet/catot*. However, this IVORN will not resolve from the global VO registry, but only from a repository that has this stream of events.
- *ivo://nvo.caltech/voeventnet/catot*

This Stream identifier can be looked up in any VO registry, returning a description, who runs it, the names, semantics, and descriptions of the parameters used in the events, how to subscribe, etc. In this case, the stream represents optical transients from the CRTS survey. For resolving the event itself, we want a repository that will have the event, so a query would be used like this: "*Find repositories that keep the events from this Stream*"

The nature of a standard service to query VOEvent server records and the metadata necessary to describe a Stream remain under discussion in the VOEvent Working Group.

### 2.3 Authentication and Authorization

VOEvents provide a mechanism for alerting members of the astronomical community to time-critical celestial phenomena. As a result of such an alert, significant hardware, software and personnel assets of the community may be retargeted to investigate those phenomena. The scientific and financial costs of such retargeting may be large, but the potential scientific gains are larger. The success of VOEvent — and of observations of astronomical transients in general — depends on minimizing both intentional and unintentional noise/spam associated with this communications channel. All of the familiar internet security worries apply to VOEvents. A discussion of these issues is available under Authentication [34] from both the VOEvent standpoint and for comparison, general XML signatures.

## 3. VOEvent Semantics

A VOEvent packet provides a general purpose mechanism for representing information about transient astronomical events. However, not all VO data are suitable for expression using VOEvent. The VOEvent schema [20] is as simple as practical to allow the minimal representation of scientifically meaningful, time critical, events. VOEvent also borrows other standard VO and astronomical schema, specifically STC for space-time coordinates. The usual IVOA standards such as registries and UCD identifiers are used. VOEvent has a strong interest in the development of complete and robust astronomical ontologies, but must rely on pragmatic and immediately useful prototypes of planned facilities.

By definition, a VOEvent packet contains a single XML **<VOEvent>** element. If multiple **<VOEvent>** elements are jointly contained within a larger document in some fashion, they should still be handled as separate alert packets. A **<VOEvent>** element may contain at most one of each of the following optional sub-elements:

| | |
|---|---|
| **<Who>** | *Identification of scientifically responsible Author* |
| **<What>** | *Event Characterization modeled by the Author* |
| **<WhereWhen>** | *Space-Time Coordinates of the event* |
| **<How>** | *Instrument Configuration* |
| **<Why>** | *Initial Scientific Assessment* |
| | |
| **<Citations>** | *Follow-up Observations* |
| **<Description>** | *Human Oriented Content* |
| **<Reference>** | *External Content* |

Only those elements required to convey the event being described need be present; the ordering of elements is not formally constrained. The intent of VOEvent is to describe a single astronomical transient event per packet. Multiple events should be expressed using multiple packets. On the other hand, complex observations may best be expressed using multiple follow-up packets or via embedded **<References>** to external resources such as VOTables or RTML documents. XML structures other than those listed in this document should be used with care within a **<VOEvent>** element, but some applications may require the freedom to reference schema outside the scope of this specification. Section 4 contains examples of complete VOEvent packets.

## 3.1 <VOEvent> — identifiers, roles and versions

A **<VOEvent>** expresses the discovery of a sky transient event, located in a region of space and time, observed by an instrument, and published by a person or institution who may have developed a hypothesis about the underlying classification of the event.

The **<VOEvent>** element has three attributes:

**3.1.1 *ivorn*** — Each VOEvent packet is required to have one-and-only-one identifier, expressed with the ***ivorn*** attribute. VOEvent identifiers are URIs [15]. As the issuance of duplicate identifiers would diminish the trust placed in systems exchanging VOEvents, it is anticipated that a number of VOEvent publishers will be founded to issue unique IVORNs from a variety of useful and appropriate namespaces. The non-opaque URI identifier is constructed systematically so that the identifier of a different resource, the VOEventStreamRegExt, is deducible from the identifier of an event. The first part is the identifier for the publisher, and the event identifier is built from this, then a # symbol, then a local string that is meaningful only in the context of that publisher.

**3.1.2 *role*** — The optional ***role*** attribute accepts the enumerated options:

- The value "*observation*" is the default if the ***role*** is missing; this means that the packet describes an observation of the actual universe.
- The value "*prediction*" indicates that the VOEvent describes an event of whatever description that has yet to occur when the packet is created.
- The value "*utility*" means that the packet expresses nothing about astrophysics, but rather information about the observing system. This could be used, for example, for a satellite to express that it has changed its configuration.
- The value "*test*" means that the packet does not describe actual astronomical events, but rather is part of a testing procedure of some kind.

It is the responsibility of all who receive VOEvent packets to pay attention to the *role*, and to be quite sure of the difference between an actual event and a test of the system or a prediction of an event that has yet to happen.

**3.1.3 *version*** — The ***version*** attribute is required to be present and to equal "2.0" for all VOEvent packets governed by this version of the standard. There is no default value.

For example, a **<VOEvent>** packet resulting from Tycho Brahe's discovery of a "Stella Nova" in Cassiopeia on 11 November 1572 [24] might start:

```
  <VOEvent ivorn="ivo://uraniborg.hven#1572-11-11/0001" role="observation" version="2.0"
xmlns:... >
```

The ***xmlns*** attribute refers to one-or-more standard XML namespace declarations that may optionally help define the contents of a packet.

## 3.2 <Who> — Curation Metadata

This element of a VOEvent packet is devoted to curation metadata: who is responsible for the information content of the packet. Usage should be compatible with section 3.2 of the IVOA Resource Metadata specification [16]. Typical curation content would include:

### 3.2.1<Author>

Author information follows the IVOA curation information schema: the organization responsible for the packet can have a title, short name or acronym, and a logo. A contact person has a name, email, and phone number. Other contributors can also be noted.

An example of Author information might be:

```
<Author>
    <title>Rapid Telescope for Optical Response</title>
    <shortName>Raptor</shortName>
    <logoURL>http://www.raptor.lanl.gov/images/RAPTOR_patchLarge.jpg</logoURL>
    <contactName>Robert White</contactName>
    <contactEmail>rwhite@lanl.gov</contactEmail>
    <contactPhone>+1 800 555 1212</contactPhone>
 </Author>
```

### 3.2.2<AuthorIVORN>

As an alternative to quoting Author information over and over, this information can be published to the VO registry, then referenced through an IVORN. The **<AuthorIVORN>** element contains the identifier of the organization responsible for making the VOEvent available. Event subscribers will often use this as their primary filtering criterion. Many subscribers will only want events from a particular publisher, or more precisely, from a specific content creator. In general, **<AuthorIVORN>** should be a VOResource identifier that resolves to an organization in the sense of [16]. Publishers and subscribers may use this VOResource to exchange curation metadata directly.

### 3.2.3 <Date>

The **<Date>** contains the date and time of the creation of the VOEvent packet. The required format is a subset of ISO-8601 (*e.g.*, *yyyy-mm-dd**T**hh:mm:ss*, see [28]). The timescale — for curation purposes only — is assumed to be Coordinated Universal Time (UTC). Discussions of date and time for the expression of meaningful scientific coordinates may be found in [17] and [26].

Minimal **<Who>** usage might resemble:

```
<Who>
     <AuthorIVORN>ivo://uraniborg.hven/Tycho</AuthorIVORN>
     <Date>1573-05-05T01:23:45Z</Date>
</Who>
```

Tycho first noted SN 1572 on 11 November of that year. The event was published in Tycho's pamphlet *De Stella Nova* by 5 May 1573, thus this later date is placed in the curation metadata. More detailed curation metadata can be retrieved directly from the publisher.

## 3.3 <What> — Event Characterization

The **<What>** and **<Why>** elements work together to characterize the nature of a VOEvent. That is: **<What>** has author-defined parameters about what was measured directly, or other relevant information about the event, versus **<Why>** is a data model of fixed schema about the hypothesized underlying cause or causes of the astrophysical event.

In general, an observation is the association of one or more dependent variables with zero or more independent variables. The **<WhereWhen>** element, for example, is often used to express the independent variables in an observation — where was the telescope pointed and when was the camera shutter opened. The **<What>** element, on the other hand, is typically used to express the dependent variables — what was seen at that location at that time.

A **<What>** element contains a list of **<Param>** elements which may be associated and labeled using **<Group>** elements. It may also have one or more **<Table>** elements, each of which can contain **<Param>** and **<Field>** elements: these last define a whole column, or vector of data, rather than a single primitive value as with **<Param>**. See §4.1 for an example of usage.

### 3.3.1 <Param> — Numbers and strings with semantics

**<Param>** elements may be used to represent the values of arbitrarily named quantities. Thus a publisher need not establish a fixed schema for all events they issue. Unified Content Descriptors (UCDs) [18] may be used to clarify meaning. Usage of **<Param>** and **<Group>** is similar to the VOTable specification, see §4.1 of [21].

A **<Param>** may contain elements **<Description>** and **<Reference>**. Like most VOEvent elements, these can be used to give further descriptive documentation about what this parameter means. The **<Param>** may also contain an element **<Value>** for the value of the parameter, as an alternate to the 'value' attribute defined below: if both are present, the attribute takes precedence over the element. This allows parameter values to include a richer variety of text strings, to avoid strings being changed by Attribute-Value normalization [30] that is part of the XML specification.

The following attributes are supported for **<Param>**:

3.3.1.1 ***name*** — A simple utilitarian name. This name may or may not have significance to subscribing clients.

3.3.1.2 ***value*** — A string representing the value in question. No range or type checking of implied numbers is performed.

3.3.1.3 ***unit*** — The unit for interpreting ***value***. See §4.3 of [21] which relies on [25].

3.3.1.4 ***ucd*** — A UCD [18] expression characterizing the nature of the **<Param>**.

3.3.1.5 ***dataType*** — A string specifying the data type of the **<Param>**. Allowed values are "string", "int", or "float", with the default being "string".

- For dataType=float, the value must contain a possibly signed decimal or floating point number as described in [36], possibly embedded in whitespace; it may also be [+|-]nan or [+|-]inf. If the value cannot be parsed this way, for example null string, it may return zero or NaN, but no exception should be thrown.
- For dataType=int, the value must contain a possibly signed decimal number, possibly embedded in whitespace. Conversion of floating point numbers to integers truncates (towards zero). If the value cannot be parsed this way, for example null string, it will return zero, and no exception should be thrown.

3.3.1.6 ***utype*** — A utype [32] defines this **<Param>** as part of a larger data structure, such as one of the IVOA standard data models.

For example, here are three values from a GCN [4] notice:

```
TRIGGER_NUM = 114299 RATE_SIGNIF = 20.49 GRB_INTEN = 73288
```

In VOEvent, these can be represented as:

```
<Param name="TRIGGER_NUM" value="114299" ucd="meta.id" />
 <Param name="RATE_SIGNIF" value="20.49"  ucd="stat.snr" dataType="float">
     <Description>Best significance after trying all algorithms</Description>
     <Reference uri="http://gcn.gsfc.nasa.gov/swift.html"/>
  </Param>
 <Param name="GRB_INTEN" value="73288"  ucd="phot.count" dataType="int"/>
```

### 3.3.2 <Group> — collection of related Params

**<Group>** provides a simple mechanism for associating several **<Param>** (and/or **<Reference>**) elements, for instance, an error with a measurement. **<Group>**s may NOT be nested. **<Group>** elements may have a ***name*** attribute, and unlike VOTable usage, may also have a ***type*** attribute:

3.3.2.1 ***name*** — A simple name such as in §3.3.1.1.

3.3.2.2 ***type*** — A string that can be used to build data structures, for example a Group with type "complex" might have Params called "real" and "imag" for the two components of a complex number.

In a GCN notice, for example, we might see this line:

```
GRB_INTEN:       73288 [cnts]    Peak=1310 [cnts/sec]
```

which could be expressed with one Param with a Value element, and the other with a Value attribute:

```
<Group type="GRB_INTEN">
     <Param name="cnts" ucd="phot.count" dataType="int">
              <Value>73288</Value>
    </Param>
    <Param name="peak" value="1310"  ucd="arith.rate;phot.count" dataType="float"/>
</Group>
```

Note also that there cannot be Groups within Groups: a Group may only contain Params and not Groups or Tables; a Table may only contain Params and Fields and not Groups or Tables. There are rules of uniqueness for Params, Groups, Fields and Tables in VOEvent:

- Each Param and Field must have a name. A Group or Table without a name is equivalent to having a name which is the null string.
- Names must be unique within the set of those Params that are not in a Group or Table.
- Names must be unique for the set of Params and Fields within a given Group or Table.
- Groups and Tables must have unique names: this means that only one Group or Table can be nameless.

### 3.3.3 <Table> — simple tabular data

This element is intended for a short and simple table, and re-uses the ideas and syntax of the IVOA VOTable, but simplified and streamlined: this is appropriate because complex tables can be written as full VOTable and linked from the VOEvent. Specifically, these simplifications are: no support for hierarchy of tables (RESOURCE); no internal references (FieldRef and ParamRef); no provision for binary data, only XML; table cells can only be string, float, or int, in place of the arrays of 12 possible types and extensions; no formatting information contained in the Table, nor domain of the data (VALUES); no referencing between cells; there is no INFO element.

There are five elements defined in this subsection: Table, Field, Data, TR, TD.

A **<Table>** element can contain a sequence of **<Field>** elements, one for each column of the table, and **<Param>** elements for scalar information about the table. There is then a single **<Data>** element that contains the data of the table, which is represented as a sequence of table rows, which are **<TR>** elements, each containing a sequence of **<TD>** elements for the table cells. For a full table, where every cell has a value, the number of **<TD>** elements in each row should be the same as the number of **<Field>** elements. There is then a 1-to-1 correspondence between them for each row.

The Table can contain **<Description>** and **<Reference>** elements to add documentation; the **<Field>** elements can also contain these. Thus the **<Table>** can contain, in order, an optional **<Description>** and **<Reference>**, then a sequence of one or more **<Field>** elements, then a **<Data>** element. The **<Field>** element can also contain optional **<Description>** and **<Reference>** and nothing else. The **<Data>** element can contain only **<TR>** elements, each of which can contain only **<TD>** elements. The following explains the attributes that are allowed for these five elements.

The following attributes are supported for **<Table>**:

3.3.3.1 ***name*** — A simple utilitarian name that may be used for identification or presentation purposes. This name may or may not have significance to subscribing clients.

3.3.3.2 ***type*** — A string representing the type of the Table, that consumers can use for presentation or parsing. For example, a table of type "spectralLines" could mean to some community to expect columns (i.e. the **<Field>**s) named "wavelength", "width", "name" to define spectral lines.

The **<Field>** element defines the semantic nature of a Table column, and is structured similarly to the **<Param>** element of section 3.3.1. The following attributes are supported for **<Field>**, similarly to the **<Param>** definition above:

3.3.3.3 ***name*** — A simple utilitarian name that may be used elsewhere in the packet. This name may or may not have significance to subscribing clients.

3.3.3.4 ***unit*** — The unit for interpreting the values as given in the **<TD>** table cells See §4.3 of [21] which relies on [25].

3.3.3.5 ***ucd*** — A UCD [18] expression characterizing the nature of the data in this table column.

3.3.3.6 ***dataType*** — A string specifying the data type of the table column. Allowed values are "string", "int", or "float", with the default being "string".

3.3.3.7 ***utype*** — A utype [31] defines this **<Param>** as part of a larger data structure, such as one of the IVOA standard data models.

The following is an example of a Table element. Note the **dataType** attribute that is used to interpret the values in the table cells.

```
    <Table>
         <Description>Individual Moduli and Distances for NGC 0931 from NED</Description>
         <Field name="(m-M)"    unit="mag" ucd="phot.mag.distMod" dataType="float"/>
         <Field name="err(m-M)" unit="mag" ucd="phot.mag.distMod;stat.err" dataType="float"/>
         <Field name="D"        unit="Mpc" ucd="pos.distance dataType="float"/>
         <Field name="REFCODE"             ucd="meta.bib.bibcode"/>
         <Data>
  <TR><TD>33.16</TD><TD>0.38</TD><TD>42.9</TD><TD>1997ApJS..109..333W</TD></TR>
  <TR><TD>33.32</TD><TD>0.38</TD><TD>46.1</TD><TD>1997ApJS..109..333W</TD></TR>
  <TR><TD>33.51</TD><TD>0.48</TD><TD>50.4</TD><TD>2009ApJS..182..474S</TD></TR>
  <TR><TD>33.55</TD><TD>0.38</TD><TD>51.3</TD><TD>1997ApJS..109..333W</TD></TR>
  <TR><TD>33.71</TD><TD>0.43</TD><TD>55.2</TD><TD>2009ApJS..182..474S</TD></TR>
  <TR><TD>34.01</TD><TD>0.80</TD><TD>63.3</TD><TD>1997ApJS..109..333W</TD></TR>
         </Data>
       </Table>
```

## 3.4 <WhereWhen> — Space-Time Coordinates

A VOEvent packet will typically include information about where in the sky and when in time an event was detected, and from what location, along with spatial and temporal coordinate systems and errors. If either the spatial or temporal locators are absent, it is to be assumed that the information is either unknown or irrelevant. VOEvent v2.0 uses the syntax of the IVOA Space-Time Coordinate (STC) specification version 1.30 or later; the **<WhereWhen>** element may reference an STC [17] **<ObsDataLocation>** element to provide a sky location and time for the event. VOEvent publishers should construct expressions that concisely provide all information that is scientifically significant to the event, and no more than that. See §4.1 for an example of usage.

STC expressions are used to locate the physical phenomena associated with a VOEvent alert in both time and space as described below. The **<ObsDataLocation>** element is a combination of information describing the location of an observation in the sky along with information describing the location of the observatory from which that observation was made. Both the sky and the observatory are in constant motion, and STC inextricably relates spatial and temporal information.

```
<WhereWhen>
     <ObsDataLocation>
         <ObservatoryLocation/>
         <ObservationLocation/>
     </ObsDataLocation>
 </WhereWhen>
```

### 3.4.1 ObservationLocation

The **<ObservationLocation>** defines the location of the event; the **<ObservatoryLocation>** specifies the location of the observatory for which that location is valid. It should contain a link to a coordinate system, **<AstroCoordSystem>**, as well as the actual coordinates of the event, **<AstroCoords>**, containing a reference back to the coordinate system specification. For example:

```
<ObservationLocation>
      <AstroCoordSystem id="UTC-FK5-GEO" />
      <AstroCoords coord_system_id="UTC-FK5-GEO">
         <Time unit="s">
             <TimeInstant>
                 <ISOTime>2004-07-15T08:23:56</ISOTime>
             </TimeInstant>
             <Error>2</Error>
         </Time>
         <Position2D unit="deg">
             <Value2>
                 <C1>148.88821</C1>
                 <C2>69.06529</C2>
             </Value2>
             <Error2Radius>0.03</Error2Radius>
         </Position2D>
     </AstroCoords>
 </ObservationLocation>
```

Specifying errors is optional but recommended for both time and space components.

The **<AstroCoords>** element has a ***coord_system_id*** attribute and the **<AstroCoordSystem>** has a ***id*** attribute. The value of both of these should be identical, and represent the space-time coordinate system that will be used for the event position and time.

A ***coord_system_id*** and ***id*** are built from a time part, a space part, and a "center" specification, concatenated in that order and separated by hyphens. Astronomical coordinate systems are extremely varied, but all VOEvent subscribers should be prepared to handle coordinates expressed as combinations of these basic defaults:

- The time part can be *UTC* (Coordinated Universal Time [26]), *TT* (Terrestrial Time, currently 65.184 seconds ahead of UTC), *GPS* time, or *TDB* (Barycentric Dynamical Time).
- The space part can be equatorial coordinates (right ascension and declination) expressed in either the *ICRS* or *FK5* coordinate systems.
- The center specification can be *TOPO* (*i.e.*, the location of the observatory), *GEO* (geocentric coordinates), or *BARY* (relative to the barycenter of the solar system).

It is assumed that the center reference position (origin) is the same for both space and time coordinates. That means, for instance, that *BARY* should only be paired with *TDB* (and vice-versa). See the STC specification [17] for further discussion. The list of **<AstroCoordSystem>** defaults that VOEvent brokers and clients may be called upon to understand is:

*TT-ICRS-TOPO, UTC-ICRS-TOPO, TT-FK5-TOPO, UTC-FK5-TOPO, GPS-ICRS-TOPO, GPS-FK5-TOPO, TT-ICRS-GEO, UTC-ICRS-GEO, TT-FK5-GEO, UTC-FK5-GEO, GPS-ICRS-GEO, GPS-FK5-GEO, TDB-ICRS-BARY, TDB-FK5-BARY*

The STC specification, in particular **<ObsDataLocation>** and its contained elements, allows more exotic coordinate systems (for example, describing planetary surfaces). Further description of how VOEvent packets might be constructed to convey such information to subscribers is outside the scope of this document. As with other elements of an alert packet, subscribers must be prepared to understand coordinates expressing the science and experimental design pertinent to the particular classes of sky transients that are of interest.

In short, subscribers are responsible for choosing what VOEvent packets and thus ***coord_system_id*** values they will accept. Further, subscribers may choose not to distinguish between coordinate systems that are only subtly different for their purposes — for instance between *ICRS* or *FK5*, or between *TOPO* or *GEO*. As software determines whether a packet's ***coord_system_id*** describes a supported coordinate system, the question is also what accuracy is required and what coordinate transformations may be implicitly or explicitly performed to that level of accuracy.

A similar question faces the authors of VOEvent packets, who must make a judicious choice between the available coordinate system options to meet the expected scientific needs of consumers of those packets. If a detailed or high accuracy coordinate system selection is not needed, ***UTC-ICRS-TOPO*** would be a good choice as an interoperability standard.

### 3.4.2 ObservatoryLocation

The **<ObservatoryLocation>** element is used to express the location from which the observation being described was made. It is a required element for expressing topocentric coordinate systems. An instance of **<ObservatoryLocation>** may take two forms. In the first, an observatory location may be taken from a library, for example:

```
<ObservatoryLocation id="Palomar" />
```

The id here indicates the name of the observatory, other examples being: Keck, KPNO, JCMT, MMTO, VLA, *etc*., or it may indicate one of the following generic observatory locations:

- *GEOSURFACE* - any location on the surface of the earth
- *GEOLEO* - any location in Low Earth Orbit (altitude<700 km)
- *GEOGSO* - any location within Geostationary orbit altitude
- *GEONBH* - any location within 50,000 km of the geocenter
- *GEOLUN* - any location within the Moon's orbit

For example, a packet might contain the following **<ObservatoryLocation>** to indicate that the coordinates expressed in the **<WhereWhen>** element are located with an accuracy comprising the Earth's surface:

```
<ObservatoryLocation id="GEOSURFACE" />
```

The second option for **<ObservatoryLocation>** is that an observatory can be located by specifying the actual coordinate values of longitude, latitude and altitude on the surface of the Earth. Note the use of a coordinate system for the surface of the Earth (UTC-GEOD-TOPO) is natural for an observatory location, whereas coordinate systems in the previous section are for astronomical events.

```
<ObservatoryLocation id="KPNO">
      <AstroCoordSystem id="UTC-GEOD-TOPO" />
      <AstroCoords coord_system_id="UTC-GEOD-TOPO">
         <Position3D>
             <Value3>
                 <C1 pos_unit="deg">248.4056</C1>
                 <C2 pos_unit="deg">31.9586</C2>
                 <C3 pos_unit="m">2158</C3>
             </Value3>
         </Position3D>
     </AstroCoords>
 </ObservatoryLocation>
```

### 3.4.3 Parsing the WhereWhen Element

When parsing a VOEvent packet, the following pseudocode may be of use to provide the time, the right ascension and the declination, if the author used *ICRS* spatial coordinates and *UTC* time.

```
Let  x =/voe:VOEvent/WhereWhen/ObsDataLocation/ObservationLocation/AstroCoords
If x[@coord_system_id='UTC-ICRS-TOPO'] then
    Let Time = x/Time/TimeInstant/ISOTime
    Let RA = x/Position2D/Value2/C1
    Let Dec = x/Position2D/Value2/C2
```

The coordinate system is first checked to verify that it is set to a specific value(s), *UTC-ICRS-TOPO*. In practice, a subscriber may not care about the difference between *ICRS* and *FK5* (of the order of 0.01") or between *TOPO* and *GEO* (in terms of timing, this is of the order of 25 ms for ground-based and low-earth-orbit observatories). Software may be written to simply accept anything that contains *ICRS* or *FK5*, *TOPO* or *GEO*.

### 3.4.4 Solar Events

The following coordinate systems are recognized for solar event data:

- *UTC-HPC-TOPO* - Cartesian helio-projective coordinates (solar disk)
- *UTC-HPR-TOPO* - Polar helio-projective coordinates (coronal events)
- *UTC-HGS-TOPO* - Stonyhurst heliographic coordinates
- *UTC-HGC-TOPO* - Carrington heliographic coordinates

What this means is that these coordinate combinations will be supported in the library and that, hence, use of VOEvent by the solar research community is supported. It does not imply, of course, that all VOEvent participants are expected to recognize and handle these solar coordinates - nor, for that matter, that solar subscribers be able to handle equatorial coordinates.

### 3.4.5 Events Observed from Spacecraft

Transient event alerts resulting from observations made on distant spacecraft may reference coordinates that require correction for ground-based follow-up. The precise definition of "distant" will depend on the objects observed, the instrumentation and the science program. For remote objects such as gamma-ray bursts or supernovae, it is likely that spatial coordinates measured from spacecraft in Earth orbit will be immediately useful — indeed, the error box of the reported coordinates may be much larger than than the pointing accuracy of the follow-up telescope. On the other hand, the field of view of the instrument on that telescope may be many times larger than the error box. Subscribers must always balance such concerns — this is just one facet of matching "scientific impedance" between discovery and follow-up observations.

Even if the spatial targeting coordinates require no correction, the light travel time may be quite significant between a spacecraft and any follow-up telescopes on the Earth. Subscribers may need to adjust wavefront arrival times to suit.

Authors of such events may choose to handle reporting the location of the spacecraft in different ways. First, they may simply construct the complex **<ObservatoryLocation>** element that correctly represents the rapidly moving location of an orbiting observatory. Further discussion of this topic is outside the scope of the present document, see the STC specification [17]. Of course, any subscribers to such an event stream would have to understand such an **<ObservatoryLocation>** in detail and be able to calculate appropriate time-varying adjustments to the coordinates in support of their particular science program.

Alternately, an author of event alert packets resulting from spacecraft observations might simply choose to correct their observations themselves into geocentric or barycentric coordinates. Finally, for spacecraft in Earth orbit, authors might choose to report an **<ObservatoryLocation>** such as *GEOLUN*, indicating a rough position precise to the width of the Moon's orbit. These two options might be combined by both making a geocentric correction — for instance, to simplify the handling of timing information — with the reporting of a *GEOLEO* location, for example.

## 3.5 <How> — Instrument Configuration

The **<How>** element supplies instrument specific information. A VOEvent describes events in the sky, not events in the focal plane of a telescope. Only specialized classes of event will benefit from providing detailed information about instrumental or experimental design. A **<How>** contains zero or more **<Reference>** elements (see §3.9) and **<Description>** elements, that together characterize the instrument(s) that produced the observation(s) that resulted in issuing the VOEvent packet. A URI pointing to a previous VOEvent asserts that an identical instrumental configuration was used:

```
<How>
    <Description> The Echelle spectrograph </Description>
    <Reference uri="http://nsa.noao.edu/kp012345.rtml" />
</How>
```

## 3.6 <Why> — Initial Scientific Assessment

**<Why>** seeks to capture the emerging concept of the nature of the astronomical objects and processes that generated the observations noted in the **<What>** element. Natural language words and phrases are used to express the hypothesized astrophysics, pending a standard VO ontology or formal UCD-like vocabulary of astronomical concepts (see [18] and [19], for example). The **<Why>** element has two optional attributes, ***importance*** and ***expires***, providing ratings of the relative noteworthiness and urgency of each VOEvent, respectively. Subscribers should consider the ***importance*** and ***expires*** ratings from a particular publisher in combination with other VOEvent metadata in interpreting an alert for their purposes. The publishers of each category of event are encouraged to develop a self-consistent rating scheme for these values.

**3.6.1** ***importance*** — The ***importance*** provides a rating of the noteworthiness of the VOEvent, expressed as a floating point number bounded between 0.0 and 1.0 (inclusive). The meaning of ***importance*** is unspecified other than that larger values are considered of generally greater importance.

**3.6.2** ***expires*** — The ***expires*** attribute provides a rating of the urgency or time-criticality of the VOEvent, expressed as an ISO-8601 [28] representation of some date and time in the future. The meaning of ***expires*** is application dependent but will often represent the date and time after which a follow-up observation might be belated.

A **<Why>** element contains one or more **<Concept>** and **<Name>** sub-elements. These may be used to assert concepts that specify a scientific classification of the nature of the event, or rather to attach the name of some specific astronomical object or feature. These may be organized using the **<Inference>** element, which permits expressing the nature of the ***relation*** of the contained elements to the event in question as well as an estimate of its likelihood via its ***probability*** attribute.

### 3.6.3 <Concept> — classification

The value of a **<Concept>** element uses a controlled vocabulary to express the hypothesized astrophysics. This standard VO ontology or formal UCD-like vocabulary of astronomical concepts vocabulary is still in development (see [18] and [19], for example).

### 3.6.4<Description> — natural language

This element provides a natural language description of the concept, either as a replacement for the **<Concept>** element, or as an elaboration.

### 3.6.5<Name> — identification

**<Name>** provides the name of a specific astronomical object. It is preferred, but not required, to use standard astronomical nomenclature, *e.g.*, as recognized by NED [22] or SIMBAD [23].

### 3.6.6<Inference> — hypotheses inferred

An **<Inference>** may be used to group or associate one or more **<Name>** or **<Concept>** elements. **<Inference>** has two optional attributes, ***probability*** and ***relation***:

3.6.6.1 ***probability*** — The ***probability*** attribute is an estimate of the likelihood of the **<Inference>** accurately describing the event in question. It is expressed as a floating point number bounded between 0.0 and 1.0 (inclusive). In particular, note that a ***probability*** of 0.0 can be used to eliminate **<Inferences>** from further consideration.

3.6.6.2 ***relation*** — The ***relation*** attribute is a natural language string that expresses the degree of connection between the **<Inference>** and the event described by the packet. Typical values might be "associated" - a SN is associated with a particular galaxy - or "identified" - a SN is identified as corresponding to a particular precursor star. Such a one-to-one identification is considered to be the default ***relation*** in the absence of the attribute.

This example asserts that the creator of the packet is 100% certain that the event being described is equivalent to *Tycho's Star*, which has been identified as a *Type Ia Supernova*, and is "associated" with the *SN remnant* known as *3C 10*. This was an important discovery, but is no longer a very urgent one:

```
<Why importance="1.0" expires="1574-05-11T12:00:00">
     <Inference probability="1.0">
              <Name>Tycho's Stella Nova</Name>
              <Concept>http://ivoat.ivoa.net/stars.supernova.Ia</Concept>
    </Inference>
    <Inference probability="1.0" relation="associated">
             <Name>3C 10</Name>
              <Concept>http://ivoat.ivoa.net/ISM.SNRemnant</Concept>
              <Description>Supernova remnant</Description>
    </Inference>
</Why>
```

## 3.7 <Citations> — Follow-up Observations

A VOEvent packet without a **<Citations>** element can be assumed to be asserting information about a new celestial discovery. Citations reference previous events to do one of three things:

- follow-up an event alert with more observations or other relevant data, or
- supersede a prior event with better, equivalent information, or
- issue a complete retraction of a previous event.

Citations form the edges of a directed graph whose nodes are VOEvent instances; they allow merging multiple events into a single related thread, a way to collect multi-sourced data into a coherent whole. Projects that implement VOEvent handling may decide to implement for different conditions of citation -- perhaps assuming a sparse or structured citation graph, or a small or large arity for each event. We recommend that the meaning of 'citation' should be a strong one: *if a reader is to understand an event, then the reader should understand the cited event*. This is the relation between a comment and a post, between one observation of a transient and another relevant observation. However, not everything should be cited: while the papers of Einstein may be relevant, they need not be always cited! A different notion is that of association of sources: as in a radio source being near an optical source. If an author wishes to express this notion, the **<Inference>** element can carry this information (see section 3.6.6).

A **<Citations>** element contains one or more **<EventIVORN>** elements. The standard does not attempt to enforce references to be logically consistent; this is the responsibility of publishers and subscribers.

### 3.7.1 <EventIVORN> — Cited event and relationship

An **<EventIVORN>** element contains the IVORN of a previously published VOEvent packet. Each **<EventIVORN>** describes the relationship of the current packet to that previous VOEvent. It has one required attribute:

3.7.1.1 ***cite*** — The ***cite*** attribute accepts three possible enumerated values, "*followup*", "*supersedes*" or "*retraction*". There is no default value.

The value of the ***cite*** attribute modifies the VOEvent semantics. In contrast to a VOEvent announcing a discovery (*i.e.*, a packet with no citations), a VOEvent may be explicitly a "*followup*", citing one or more earlier packets — meaning that the described real or virtual observation was done as a response to those cited packet(s). In this case, the supplied information is assumed to be a new, independent measurement.

The ***cite*** may be "*supersedes*", which can be used to express a variety of possible event contingencies. A prior VOEvent may be superseded, for example, if reprocessing of the original observation has resulted in different values for quantities expressed by **<What>** or **<WhereWhen>** or if the investigators have formed a new **<Why>** regarding the event. On the other hand, if a later observation has simply resulted in different measurements to report, this would typically be issued as a "*followup*".

When a citation is made with a "supersedes" or "retraction" attribute, it is assumed that *all* of the previous information is superseded: and so the cited event is no longer needed other than for archival or historical purposes. If there is datum X and datum Y in the original, and X gets improved calibration, then Y must also be copied to the new event, or else its value will no longer be seen. There is, however, no guarantee that a superseded or retracted event will not be subsequently cited or referenced.

A "*supersedes*" ***cite*** can also be used to merge two or more earlier VOEvent threads that are later determined to be related in some fashion. The VOEvents to be merged are indicated with separate **<EventIVORN>** elements. The proper interpretation of such a merger would depend on a VOEvent client having received all intervening packets from all relevant threads. Finally, "*supersedes*" can be used in combination with a "*followup*" to divide a single VOEvent into two or more new threads. First, follow-up the event in one packet and then supersede the original event, rather than the follow-up, in a second packet (with a second identifier that can start a second thread).

The "*retraction*" ***cite*** indicates that the initial discovery event is being completely retracted for some reason. The publisher of a retraction may be other than the publisher of the original VOEvent — subscribers are free to interpret such a situation as they see fit.

Splitting, merging or retracting a VOEvent should typically be accompanied by a **<Description>** element discussing why such actions are being taken.

An attempt is made to retract the sighting of Tycho's supernova:

```
<Citations>
    <EventIVORN cite="retraction">ivo://uraniborg.hven#1572-11-11/0001</EventIVORN>
    <Description>Oops!</Description>
</Citations>
```

## 3.8 <Description> — Human Oriented Content

A **<Description>** may be included within any element or sub-element of a VOEvent to add human readable content. **<Descriptions>** may NOT contain **<References>**. Users may wish to embellish Description sections with HTML tags such as images and URL links, and these should not be seen by the XML parser, as they will cause the VOEvent XML to be invalid against the schema. However, it is possible to use the CDATA mechanism of XML to quote text at length, so this may be used for complicated tagged Descriptions. See the example in section 4 for usage.

## 3.9 <Reference> — External Content

A **<Reference>** may be included in any element or sub-element of a VOEvent packet to describe an association with external content via a Uniform Resource Identifier [15]. In addition to the locator for the content, there is also a locator for the meaning of the content, which is another URI, specified by the ***meaning*** attribute. It is anticipated that a Note will be written discussing the IVOA-wide usage of such meaning locators. A client application may ignore **<Reference>** elements with unrecognized ***meaning*** attributes. On the other hand, the client may ignore the 'meaning' attribute if the position of the **<Reference>** element is sufficient to establish semantics; for example if it is contained in a **<Param>**, then presumably it gives drill-down semantics for the precise meaning of that **<Param>**. A **<Reference>** must be expressed as an empty element, with attributes only.

A **<Reference>** element has the attributes:

- **3.9.1 *uri*** — The identifier of another document (*anyURI* [29]). This attribute must be present.
- **3.9.2 *meaning*** — The nature of the document referenced (*anyURI*). This attribute is optional.
- **3.9.3 *mimetype*** — An optional MIME type [35] for the referenced document.
- **3.9.4 *type*** *[DEPRECATED] — The type of the document as described in VOEvent v1.11.*
- **3.9.5 *name*** *[DEPRECATED] — A short name as described in VOEvent v1.11.*

A **<Reference>** is used to provide general purpose ancillary data with well-defined meaning. Here a fits image is presented (h.fits), and also a link to the data model that is needed for a machine to understand the meaning.

```
<Group type="MyFilterWithImage">
     <Reference uri=http://.../data/h.fits
       meaning="http://www.ivoa.net/rdf/IVOAT#Filter/h"/>
</Group>
```

An example of the indirection of a VOEvent packet using **<Reference>**

```
<VOEvent ivorn="ivo://raptor.lanl#235649409/sn2005k" role="observation" version="2.0">
       <Reference uri="http://raptor.lanl.gov/documents/event233.xml"/>
</VOEvent>
```

## 4: VOEvent Example

This imaginary event is a brightness measurement of a past supernova from the RAPTOR [10] telescope. The **<What>** section reports a **<Description>** and **<Reference>** followed by a **<Param>** about seeing and a **<Group>** with the actual report: the magnitude is 19.5, measured 278.02 days after the reference time, which is reported in the **<WhereWhen>** section. There is a **<Table>** of measured distances to the presumed host galaxy. The packet represents a follow-up observation of an earlier event, as defined in the **<Citations>** element.

```
<?xml version="1.0" encoding="UTF-8"?>
<voe:VOEvent ivorn="ivo://raptor.lanl/VOEvent#235649409" role="observation" version="2.0"
  xmlns:xsi="http://www.w3.org/2001/XMLSchema-instance"
  xmlns:voe="http://www.ivoa.net/xml/VOEvent/v2.0"
  xsi:schemaLocation="http://www.ivoa.net/xml/VOEvent/v2.0
http://www.ivoa.net/xml/VOEvent/VOEvent-v2.0.xsd">
  <Who>
    <AuthorIVORN>ivo://raptor.lanl/organization</AuthorIVORN>
    <Date>2005-04-15T14:34:16</Date>
  </Who>
  <What>
    <Description>An imaginary event report about SN 2009lw.</Description>
    <Reference uri="http://raptor.lanl.gov/data/lightcurves/235649409"
      mimetype="application/x-votable+xml"
meaning="http://www.ivoa.net/rdf/IVOAT#LightCurve"/>
    <Param name="seeing" value="2" unit="arcsec" ucd="instr.obsty.site.seeing"
dataType="float"/>
    <Group name="magnitude">
      <Description>Time is days since the ref time in the WhereWhen section</Description>
      <Param name="time" value="278.02" unit="d" ucd="time.epoch" dataType="float"/>
      <Param name="mag" value="19.5" unit="mag" ucd="phot.mag" dataType="float"/>
      <Param name="magerr" value="0.14" unit="mag" ucd="phot.mag; stat.err"
dataType="float"/>
    </Group>
    <Table>
      <Param name="telescope" value="various" utype="whatever"/>
      <Description>Individual Moduli and Distances for NGC 0931 from NED</Description>
      <Field name="(m-M)" unit="mag" ucd="phot.mag.distMod"/>
      <Field name="err(m-M)" unit="mag" ucd="phot.mag.distMod;stat.err"/>
      <Field name="D" unit="Mpc" ucd="pos.distance"/>
      <Field name="REFCODE" ucd="meta.bib.bibcode" utype="whatever"/>
      <Data>
        <TR><TD>33.16</TD><TD>0.38</TD><TD>51.3</TD><TD>1997ApJS..109..333W</TD></TR>
        <TR><TD>33.32</TD><TD>0.38</TD><TD>46.1</TD><TD>1997ApJS..109..333W</TD></TR>
        <TR><TD>33.51</TD><TD>0.48</TD><TD>50.4</TD><TD>2009ApJS..182..474S</TD></TR>
        <TR><TD>33.55</TD><TD>0.38</TD><TD>51.3</TD><TD>1997ApJS..109..333W</TD></TR>
        <TR><TD>33.71</TD><TD>0.43</TD><TD>55.2</TD><TD>2009ApJS..182..474S</TD></TR>
        <TR><TD>34.01</TD><TD>0.80</TD><TD>63.3</TD><TD>1997ApJS..109..333W</TD></TR>
      </Data>
    </Table>
  </What>
  <WhereWhen id="Raptor-2455100">
    <ObsDataLocation>
      <ObservatoryLocation id="RAPTOR"/>
      <ObservationLocation>
        <AstroCoordSystem id="UTC-ICRS-TOPO"/>
        <AstroCoords coord_system_id="UTC-ICRS-TOPO">
          <Time>
            <TimeInstant>
              <ISOTime>2009-09-25T12:00:00</ISOTime>
            </TimeInstant>
            <Error>0.0</Error>
          </Time>
          <Position2D unit="deg">
            <Value2>
              <C1>37.0603169</C1>
              <!-- RA  -->
              <C2>31.3116578</C2>
              <!-- Dec -->
            </Value2>
            <Error2Radius>0.03</Error2Radius>
          </Position2D>
        </AstroCoords>
      </ObservationLocation>
    </ObsDataLocation>
  </WhereWhen>
  <How>
```

```
    <Description>
      <![CDATA[This VOEvent packet resulted from observations made with <a
href=http://www.raptor.lanl.gov>Raptor</a> AB at Los Alamos. ]]>
    </Description>
  </How>
  <Citations>
    <EventIVORN cite="followup">ivo://raptor.lanl/VOEvent#235649408</EventIVORN>
  </Citations>
  <Why>
    <Concept>http://ivoat.ivoa.net/process.variation.burst;em.opt</Concept>
    <Description>Looks like a SN</Description>
    <Inference relation="associated" probability="0.99">
      <Name>NGC0931</Name>
    </Inference>
  </Why>
</voe:VOEvent>
```

## 5. Schema Diagram for VOEvent

This image summarizes the basic structure of the event packet. The image shows how the <Description> and <Reference> elements can appear in many different places, abbreviated by D,R. Elements and their hierarchy are in black, attributes in green, required attributes underlined..

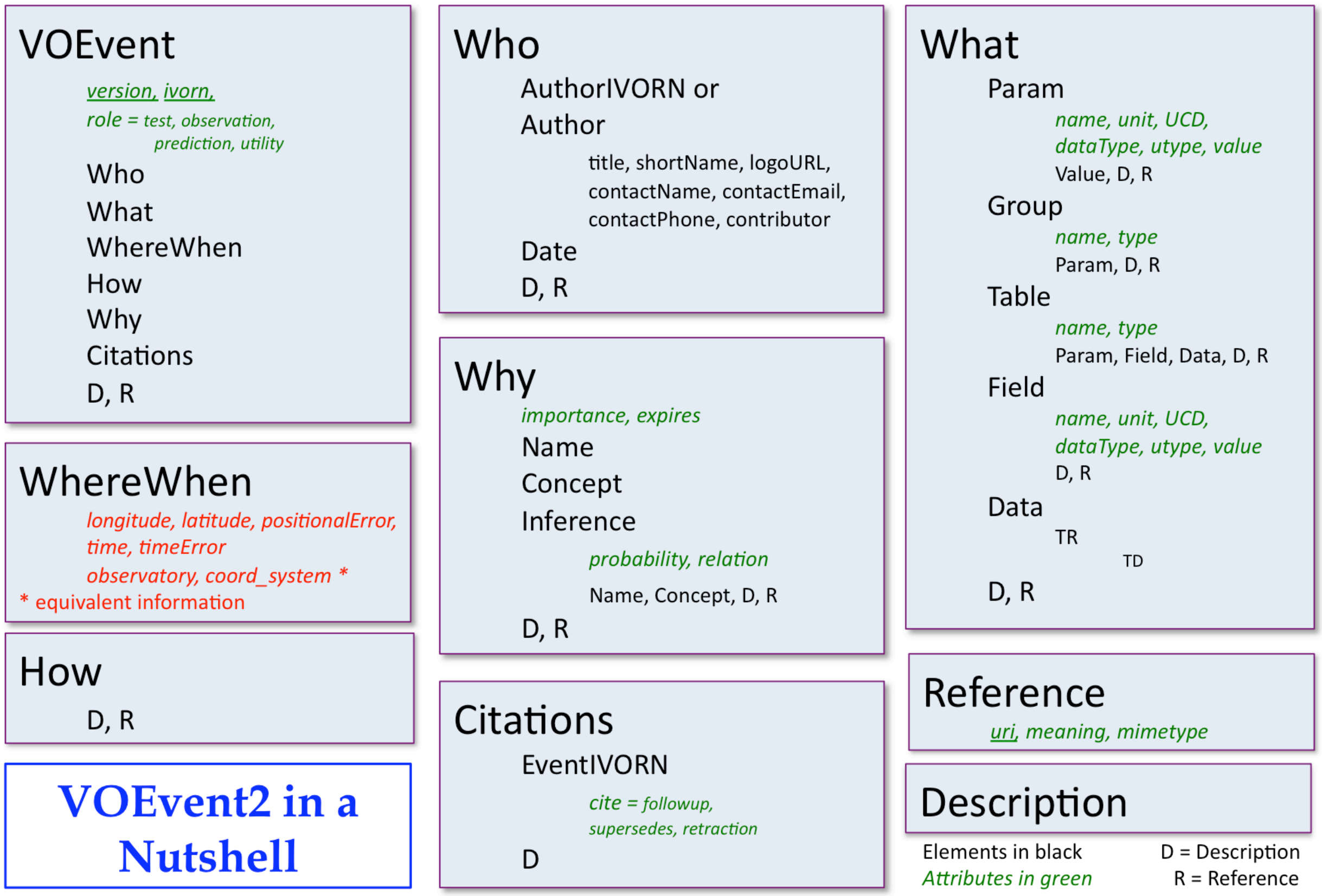

## 6. References

1. ATEL: The Astronomer's Telegram http://www.astronomerstelegram.org
2. CBAT: Central Bureau for Astronomical Telegrams
   http://cfa-www.harvard.edu/iau/cbat.html, or
   http://cfa-www.harvard.edu/iau/DiscoveryInfo.html (discovery schema)
3. eSTAR: eScience Telescopes for Astronomical Research http://www.estar.org.uk
4. GCN: The Gamma-Ray Burst Coordinates Network http://gcn.gsfc.nasa.gov
5. Skyalert: Event publisher and disseminator http://skyalert.org/
6. LIGO: Laser Interferometer Gravitational Wave Observatory http://www.ligo.caltech.edu
7. LSST: Large Synoptic Survey Telescope http://www.lsst.org
8. CRTS: Catalina Realtime Transient Survey http://crts.caltech.edu/
9. Pan-STARRS: the Panoramic Survey Telescope & Rapid Response System
   http://pan-starrs.ifa.hawaii.edu/public/index.html
10. RAPTOR: RAPid Telescopes for Optical Response
    http://www.raptor.lanl.gov, and http://www.thinkingtelescopes.lanl.gov (Thinking Telescopes Project)
11. Swift and Fermi: NASA Gamma-Ray observatories
    http://swift.gsfc.nasa.gov/docs/swift/swiftsc.html and http://fermi.gsfc.nasa.gov/
12. RoboNet: RoboNet-1.0 http://www.astro.livjm.ac.uk/RoboNet
13. ROBOT: A list of robotic telescope projects
    http://www.astro.physik.uni-goettingen.de/~hessman/MONET/links.html
14. RTML: Remote Telescope Markup Language
    http://www.uni-sw.gwdg.de/~hessman/RTML, or http://monet.uni-goettingen.de/twiki/bin/view/RTML
15. ID: IVOA Identifiers http://www.ivoa.net/Documents/latest/IDs.html
16. RM: Resource Metadata for the Virtual Observatory http://www.ivoa.net/Documents/latest/RM.html
17. STC: Space-Time Coordinates Metadata for the Virtual Observatory
    http://www.ivoa.net/Documents/latest/STC.html
18. UCD: Unified Content Descriptor
    http://www.ivoa.net/Documents/latest/UCD.html, or http://cdsweb.u-strasbg.fr/UCD
19. VOConcepts: a proposed UCD for Astronomical Objects, Events, and Processes
    http://monet.uni-sw.gwdg.de/twiki/bin/view/VOEvent/UnifiedContentDescriptors
20. VOEvent: Sky Event Reporting Metadata
    http://www.ivoa.net/Documents/latest/VOEvent.html
    http://www.ivoa.net/xml/VOEvent/VOEvent-v2.0.xsd (schema)
    http://www.ivoa.net/Documents/Notes/VOEventTransport (transport)
    http://www.ivoa.net/twiki/bin/view/IVOA/IvoaVOEvent (working group)
21. VOTable: Format Definition http://www.ivoa.net/Documents/latest/VOT.html
22. NED: NASA/IPAC Extragalactic Database http://nedwww.ipac.caltech.edu/
23. SIMBAD: Set of Identifications, Measurements and Bibliography for Astronomical Data
    http://simbad.u-strasbg.fr/Simbad
24. TYCHO: De Stella Nova http://www.texts.dnlb.dk/DeNovaStella/Index.html (in Danish)
25. UNITS: Standards for Astronomical Catalogues: Units http://vizier.u-strasbg.fr/doc/catstd-3.2.htx
26. UTC: the future of Coordinated Universal Time http://www.ucolick.org/~sla/leapsecs
27. Checksum: FITS Checksum Proposal http://fits.gsfc.nasa.gov/registry/checksum.html
28. ISO 8601: standard representation of dates and times
    http://www.w3.org/TR/NOTE-datetime.html or http://www.cl.cam.ac.uk/~mgk25/iso-time.html
29. XML: Extensible Markup Language
    http://xml.coverpages.org/xml.html, and http://xml.coverpages.org/schemas.html, or
    http://www.ucc.ie/xml (FAQ) or http://www.w3.org/TR/xmlschema-2 (Datatypes)
30. Attribute-Value Normalization http://www.w3.org/TR/REC-xml/#AVNormalize
31. PTF: Palomar Transient Factory http://www.astro.caltech.edu/ptf/
32. Utypes definition and syntax http://www.ivoa.net/cgi-bin/twiki/bin/view/IVOA/Utypes

33. Transport: VOEvent is transport neutral
http://www.ivoa.net/Documents/Notes/VOEventTransport and
http://www.ivoa.net/Documents/Notes/DakotaBroker
34. Authentication: VOEvent relies on external authentication
http://www.ivoa.net/Documents/latest/VOEventDigiSig.html and http://www.w3.org/TR/xmldsig-core
35. MIME type: http://en.wikipedia.org/wiki/Internet_media_type
36. Based on Python, floating point literals:
http://docs.python.org/reference/lexical_analysis.html#floating-point-literals

# 7. Schema

This XML schema corresponds to this document, but it is the document that is normative. It is available at **http://www.ivoa.net/xml/VOEvent/VOEvent-v2.0.xsd** .

```
<?xml version="1.0" encoding="UTF-8"?>
<xs:schema xmlns="http://www.ivoa.net/xml/VOEvent/v2.0"
xmlns:xs="http://www.w3.org/2001/XMLSchema"
  targetNamespace="http://www.ivoa.net/xml/VOEvent/v2.0"
elementFormDefault="unqualified">

  <xs:element name="VOEvent">
    <xs:annotation>
      <xs:documentation> VOEvent is the root element for describing observations of
immediate
        astronomical events. For more information, see
        http://www.ivoa.net/twiki/bin/view/IVOA/IvoaVOEvent. The event consists of at
most one of
        each of: Who, What, WhereWhen, How, Why, Citations, Description, and
        Reference.</xs:documentation>
    </xs:annotation>
    <xs:complexType>
      <!-- Can have zero or one of each of these, in any order -->
      <xs:all>
        <xs:element name="Who" type="Who" minOccurs="0" maxOccurs="1"/>
        <xs:element name="What" type="What" minOccurs="0" maxOccurs="1"/>
        <xs:element name="WhereWhen" type="WhereWhen" minOccurs="0" maxOccurs="1"/>
        <xs:element name="How" type="How" minOccurs="0" maxOccurs="1"/>
        <xs:element name="Why" type="Why" minOccurs="0" maxOccurs="1"/>
        <xs:element name="Citations" type="Citations" minOccurs="0" maxOccurs="1"/>
        <xs:element name="Description" type="xs:string" minOccurs="0" maxOccurs="1"/>
        <xs:element name="Reference" type="Reference" minOccurs="0" maxOccurs="1"/>
      </xs:all>
      <xs:attribute name="version" type="xs:token" fixed="2.0" use="required"/>
      <xs:attribute name="ivorn" type="xs:anyURI" use="required"/>
      <xs:attribute name="role" type="roleValues" default="observation"/>
    </xs:complexType>
  </xs:element>
  <xs:simpleType name="roleValues">
    <xs:restriction base="xs:string">
      <xs:enumeration value="observation"/>
      <xs:enumeration value="prediction"/>
      <xs:enumeration value="utility"/>
      <xs:enumeration value="test"/>
    </xs:restriction>
  </xs:simpleType>


  <xs:complexType name="Who">
    <xs:annotation>
      <xs:documentation> Who: Curation Metadata </xs:documentation>
    </xs:annotation>
    <xs:all>
      <!-- Can have zero or one of each of these. Schema is loose because
        should have AuthorIVORN *or* Author -->
      <xs:element name="AuthorIVORN" type="xs:anyURI" minOccurs="0"/>
      <xs:element name="Date" type="xs:dateTime" minOccurs="0"/>
      <xs:element name="Description" type="xs:string" minOccurs="0"/>
      <xs:element name="Reference" type="Reference" minOccurs="0"/>
      <xs:element name="Author" minOccurs="0">
        <xs:annotation>
          <xs:documentation> Author information follows the IVOA curation information
schema: the
            organization responsible for the packet can have a title, short name or
acronym, and a
```

```
            logo. A contact person has a name, email, and phone number. Other
contributors can also
            be noted. </xs:documentation>
        </xs:annotation>
        <xs:complexType>
          <xs:choice maxOccurs="unbounded">
            <xs:element name="title" type="xs:string"/>
            <xs:element name="shortName" type="xs:string"/>
            <xs:element name="logoURL" type="xs:anyURI"/>
            <xs:element name="contactName" type="xs:string"/>
            <xs:element name="contactEmail" type="xs:string"/>
            <xs:element name="contactPhone" type="xs:string"/>
            <xs:element name="contributor" type="xs:string"/>
          </xs:choice>
        </xs:complexType>
      </xs:element>
    </xs:all>
  </xs:complexType>


  <xs:complexType name="What">
    <xs:annotation>
      <xs:documentation> What: Event Characterization. This is the part of the data model
that is
        chosen by the Authoer of the event rather than the IVOA. There can be Params,
that may be in
        Groups, and Tables, and simpleTimeSeries. There can also be Description and
Reference as
        with most VOEvent elements. </xs:documentation>
    </xs:annotation>
    <xs:choice maxOccurs="unbounded">
      <!-- can have any number of any of these in any order  -->
      <xs:element name="Param" type="Param" minOccurs="0" maxOccurs="unbounded"/>
      <xs:element name="Group" type="Group" minOccurs="0" maxOccurs="unbounded"/>
      <xs:element name="Table" type="Table" minOccurs="0" maxOccurs="unbounded"/>
      <!--<xs:element ref="sts:SimpleTimeseries"          minOccurs="0"
maxOccurs="unbounded"/>-->
      <xs:element name="Description" type="xs:string" minOccurs="0"
maxOccurs="unbounded"/>
      <xs:element name="Reference" type="Reference" minOccurs="0" maxOccurs="unbounded"/>
    </xs:choice>
  </xs:complexType>


  <xs:complexType name="Param">
    <xs:annotation>
      <xs:documentation> What/Param definition. A Param has name, value, ucd, unit,
dataType; and
        may have Description and Reference.</xs:documentation>
    </xs:annotation>
    <xs:choice maxOccurs="unbounded">
      <xs:element name="Description" type="xs:string" minOccurs="0"/>
      <xs:element name="Reference" type="Reference" minOccurs="0" maxOccurs="unbounded"/>
      <xs:element name="Value" type="xs:string" minOccurs="0" maxOccurs="1"/>
    </xs:choice>
    <xs:attribute name="name" type="xs:string"/>
    <xs:attribute name="ucd" type="xs:string"/>
    <xs:attribute name="value" type="xs:string"/>
    <xs:attribute name="unit" type="xs:string"/>
    <xs:attribute name="dataType" type="dataType" default="string"/>
    <xs:attribute name="utype" type="xs:string"/>
  </xs:complexType>
  <xs:simpleType name="dataType">
    <xs:restriction base="xs:string">
      <xs:enumeration value="string"/>
      <xs:enumeration value="float"/>
      <xs:enumeration value="int"/>
    </xs:restriction>
  </xs:simpleType>

  <xs:complexType name="Group">
    <xs:annotation>
      <xs:documentation> What/Group definition: A group is a collection of Params, with
name and
        type attributes.</xs:documentation>
    </xs:annotation>
    <xs:choice maxOccurs="unbounded">
      <xs:element name="Param" type="Param" maxOccurs="unbounded"/>
      <xs:element name="Description" type="xs:string" minOccurs="0"/>
      <xs:element name="Reference" type="Reference" minOccurs="0"/>
    </xs:choice>
```

```
    <xs:attribute name="name" type="xs:string"/>
    <xs:attribute name="type" type="xs:string"/>
  </xs:complexType>


  <xs:complexType name="Table">
    <xs:annotation>
      <xs:documentation> What/Table definition. This small Table has Fields for the
column
        definitions, and Data to hold the table data, with TR for row and TD for value of
a table
        cell.</xs:documentation>
    </xs:annotation>
    <xs:choice maxOccurs="unbounded">
      <xs:element name="Description" type="xs:string" minOccurs="0"/>
      <xs:element name="Reference" type="Reference" minOccurs="0"/>
      <xs:element name="Param" type="Param" minOccurs="0" maxOccurs="unbounded"/>
      <xs:element name="Field" type="Field" minOccurs="0" maxOccurs="unbounded"/>
      <xs:element name="Data" type="Data" minOccurs="1" maxOccurs="1"/>
    </xs:choice>
    <xs:attribute name="name" type="xs:string"/>
    <xs:attribute name="type" type="xs:string"/>
  </xs:complexType>
  <xs:complexType name="Field">
    <xs:choice maxOccurs="unbounded">
      <xs:element name="Description" type="xs:string" minOccurs="0"/>
      <xs:element name="Reference" type="Reference" minOccurs="0"/>
    </xs:choice>
    <xs:attribute name="name" type="xs:string"/>
    <xs:attribute name="ucd" type="xs:string"/>
    <xs:attribute name="unit" type="xs:string"/>
    <xs:attribute name="dataType" type="dataType" default="string"/>
    <xs:attribute name="utype" type="xs:string"/>
  </xs:complexType>
  <xs:complexType name="Data">
    <xs:choice maxOccurs="unbounded">
      <xs:element name="TR" type="TR"/>
    </xs:choice>
  </xs:complexType>
  <xs:complexType name="TR">
    <xs:choice maxOccurs="unbounded">
      <xs:element name="TD" type="xs:string"/>
    </xs:choice>
  </xs:complexType>


  <xs:complexType name="WhereWhen">
    <xs:annotation>
      <xs:documentation> WhereWhen: Space-Time Coordinates. Lots and lots of elements
here, but the
        import is that each event has these: observatory, coord_system, time, timeError,
longitude,
        latitude, posError.</xs:documentation>
    </xs:annotation>
    <xs:choice maxOccurs="unbounded">
      <xs:element name="ObsDataLocation" type="ObsDataLocation" minOccurs="1"
maxOccurs="1"/>
      <xs:element name="Description" type="xs:string" minOccurs="0"/>
      <xs:element name="Reference" type="Reference" minOccurs="0"/>
    </xs:choice>
    <xs:attribute name="id" type="xs:ID" use="optional"/>
  </xs:complexType>

  <xs:complexType name="ObsDataLocation">
    <xs:annotation>
      <xs:documentation> Part of WhereWhen</xs:documentation>
    </xs:annotation>
    <xs:all>
      <xs:element name="ObservatoryLocation" minOccurs="1" maxOccurs="1"
type="ObservatoryLocation"/>
      <xs:element name="ObservationLocation" minOccurs="1" maxOccurs="1"
type="ObservationLocation"
      />
    </xs:all>
  </xs:complexType>

  <xs:complexType name="ObservationLocation">
    <xs:annotation>
      <xs:documentation> Part of WhereWhen</xs:documentation>
    </xs:annotation>
    <xs:all>
```

```
      <xs:element name="AstroCoordSystem" type="AstroCoordSystem" minOccurs="1"
maxOccurs="1"/>
      <xs:element name="AstroCoords" type="AstroCoords" minOccurs="1" maxOccurs="1"/>
    </xs:all>
  </xs:complexType>

  <xs:complexType name="AstroCoordSystem">
    <xs:annotation>
      <xs:documentation> Part of WhereWhen</xs:documentation>
    </xs:annotation>
    <!-- The empty sequence closes this to content -->
    <xs:sequence/>
    <xs:attribute name="id" type="idValues"/>
  </xs:complexType>
  <xs:simpleType name="idValues">
    <xs:restriction base="xs:string">
      <xs:enumeration value="TT-ICRS-TOPO"/>
      <xs:enumeration value="UTC-ICRS-TOPO"/>
      <xs:enumeration value="TT-FK5-TOPO"/>
      <xs:enumeration value="UTC-FK5-TOPO"/>
      <xs:enumeration value="GPS-ICRS-TOPO"/>
      <xs:enumeration value="GPS-ICRS-TOPO"/>
      <xs:enumeration value="GPS-FK5-TOPO"/>
      <xs:enumeration value="GPS-FK5-TOPO"/>
      <xs:enumeration value="TT-ICRS-GEO"/>
      <xs:enumeration value="UTC-ICRS-GEO"/>
      <xs:enumeration value="TT-FK5-GEO"/>
      <xs:enumeration value="UTC-FK5-GEO"/>
      <xs:enumeration value="GPS-ICRS-GEO"/>
      <xs:enumeration value="GPS-ICRS-GEO"/>
      <xs:enumeration value="TDB-ICRS-BARY"/>
      <xs:enumeration value="TDB-FK5-BARY"/>
      <!-- this one for ObservatoryLocation -->
      <xs:enumeration value="UTC-GEOD-TOPO"/>
    </xs:restriction>
  </xs:simpleType>

  <xs:complexType name="AstroCoords">
    <xs:annotation>
      <xs:documentation> Part of WhereWhen</xs:documentation>
    </xs:annotation>
    <xs:all>
      <xs:element name="Time" type="Time" minOccurs="0"/>
      <xs:element name="Position2D" type="Position2D" minOccurs="0"/>
      <xs:element name="Position3D" type="Position3D" minOccurs="0"/>
    </xs:all>
    <xs:attribute name="coord_system_id" type="idValues"/>
  </xs:complexType>

  <xs:complexType name="Time">
    <xs:annotation>
      <xs:documentation> Part of WhereWhen</xs:documentation>
    </xs:annotation>
    <xs:choice maxOccurs="unbounded">
      <xs:element name="TimeInstant" type="TimeInstant"/>
      <xs:element name="Error" type="xs:float" minOccurs="0"/>
    </xs:choice>
    <xs:attribute name="unit" type="xs:string"/>
  </xs:complexType>

  <xs:complexType name="TimeInstant">
    <xs:annotation>
      <xs:documentation> Part of WhereWhen</xs:documentation>
    </xs:annotation>
    <xs:choice maxOccurs="unbounded">
      <xs:element name="ISOTime" type="xs:string" minOccurs="0" maxOccurs="1"/>
      <xs:element name="TimeOffset" type="xs:float" minOccurs="0" maxOccurs="1"/>
      <xs:element name="TimeScale" type="xs:string" minOccurs="0" maxOccurs="1"/>
    </xs:choice>
  </xs:complexType>

  <xs:complexType name="Position2D">
    <xs:annotation>
      <xs:documentation> Part of WhereWhen</xs:documentation>
    </xs:annotation>
    <xs:all>
      <xs:element name="Name1" type="xs:string" minOccurs="0"/>
      <xs:element name="Name2" type="xs:string" minOccurs="0"/>
      <xs:element name="Value2" type="Value2"/>
      <xs:element name="Error2Radius" type="xs:float"/>
    </xs:all>
```

```
    <xs:attribute name="unit" type="xs:string"/>
  </xs:complexType>

  <xs:complexType name="Position3D">
    <xs:annotation>
      <xs:documentation> Part of WhereWhen</xs:documentation>
    </xs:annotation>
    <xs:all>
      <xs:element name="Name1" type="xs:string" minOccurs="0"/>
      <xs:element name="Name2" type="xs:string" minOccurs="0"/>
      <xs:element name="Name3" type="xs:string" minOccurs="0"/>
      <xs:element name="Value3" type="Value3"/>
    </xs:all>
    <xs:attribute name="unit" type="xs:string"/>
  </xs:complexType>

  <xs:complexType name="Value2">
    <xs:annotation>
      <xs:documentation> Part of WhereWhen</xs:documentation>
    </xs:annotation>
    <xs:all>
      <xs:element name="C1" type="C1"/>
      <xs:element name="C2" type="C2"/>
    </xs:all>
  </xs:complexType>

  <xs:complexType name="Value3">
    <xs:annotation>
      <xs:documentation> Part of WhereWhen</xs:documentation>
    </xs:annotation>
    <xs:all>
      <xs:element name="C1" type="C1"/>
      <xs:element name="C2" type="C2"/>
      <xs:element name="C3" type="C3"/>
    </xs:all>
  </xs:complexType>


  <xs:complexType name="C1">
    <xs:simpleContent>
      <xs:extension base="xs:float">
        <xs:attribute name="pos_unit" type="xs:string"/>
      </xs:extension>
    </xs:simpleContent>
  </xs:complexType>
  <xs:complexType name="C2">
    <xs:simpleContent>
      <xs:extension base="xs:float">
        <xs:attribute name="pos_unit" type="xs:string"/>
      </xs:extension>
    </xs:simpleContent>
  </xs:complexType>
  <xs:complexType name="C3">
    <xs:simpleContent>
      <xs:extension base="xs:float">
        <xs:attribute name="pos_unit" type="xs:string"/>
      </xs:extension>
    </xs:simpleContent>
  </xs:complexType>

  <xs:complexType name="ObservatoryLocation">
    <xs:annotation>
      <xs:documentation> Part of WhereWhen</xs:documentation>
    </xs:annotation>
    <xs:all>
      <xs:element name="AstroCoordSystem" type="AstroCoordSystem" minOccurs="0"
maxOccurs="1"/>
      <xs:element name="AstroCoords" type="AstroCoords" minOccurs="0" maxOccurs="1"/>
    </xs:all>
    <xs:attribute name="id" type="xs:string"/>
  </xs:complexType>

  <xs:complexType name="How">
    <xs:annotation>
      <xs:documentation> How: Instrument Configuration. Built with some Description and
Reference
        elements. </xs:documentation>
    </xs:annotation>
    <xs:choice maxOccurs="unbounded">
      <xs:element name="Description" type="xs:string"/>
      <xs:element name="Reference" type="Reference"/>
```

```
    </xs:choice>
  </xs:complexType>

  <xs:complexType name="Why">
    <xs:annotation>
      <xs:documentation> Why: Initial Scientific Assessment. Can make simple
Concept/Name/Desc/Ref
        for the inference or use multiple Inference containers for more semantic
sophistication.
      </xs:documentation>
    </xs:annotation>
    <xs:choice maxOccurs="unbounded">
      <xs:element name="Name" type="xs:string"/>
      <xs:element name="Concept" type="xs:string"/>
      <xs:element name="Inference" type="Inference"/>
      <xs:element name="Description" type="xs:string"/>
      <xs:element name="Reference" type="Reference"/>
    </xs:choice>
    <xs:attribute name="importance" type="xs:float"/>
    <xs:attribute name="expires" type="xs:dateTime"/>
  </xs:complexType>

  <xs:complexType name="Inference">
    <xs:annotation>
      <xs:documentation> Why/Inference: A container for a more nuanced expression,
including
        relationships and probability. </xs:documentation>
    </xs:annotation>
    <xs:choice maxOccurs="unbounded">
      <xs:element name="Name" type="xs:string"/>
      <xs:element name="Concept" type="xs:string"/>
      <xs:element name="Description" type="xs:string"/>
      <xs:element name="Reference" type="Reference"/>
    </xs:choice>
    <xs:attribute name="probability" type="smallFloat"/>
    <xs:attribute name="relation" type="xs:string"/>
  </xs:complexType>
  <xs:simpleType name="smallFloat">
    <xs:restriction base="xs:float">
      <xs:minInclusive value="0.0"/>
      <xs:maxInclusive value="1.0"/>
    </xs:restriction>
  </xs:simpleType>


  <xs:complexType name="Citations">
    <xs:annotation>
      <xs:documentation> Citations: Follow-up Observations. This section is a sequence of
EventIVORN
        elements, each of which has the IVORN of a cited event. </xs:documentation>
    </xs:annotation>
    <xs:sequence>
      <xs:element name="EventIVORN" type="EventIVORN" maxOccurs="unbounded"/>
      <xs:element name="Description" type="xs:string" minOccurs="0"/>
    </xs:sequence>
  </xs:complexType>

  <xs:complexType name="EventIVORN">
    <xs:annotation>
      <xs:documentation> Citations/EventIVORN. The value is the IVORN of the cited event,
the 'cite'
        attribute is the nature of that relationship, choosing from 'followup',
'supersedes', or
        'retraction'.</xs:documentation>
    </xs:annotation>
    <xs:simpleContent>
      <xs:extension base="xs:string">
        <xs:attribute name="cite" type="citeValues"/>
      </xs:extension>
    </xs:simpleContent>
  </xs:complexType>

  <xs:simpleType name="citeValues">
    <xs:restriction base="xs:string">
      <xs:enumeration value="followup"/>
      <xs:enumeration value="supersedes"/>
      <xs:enumeration value="retraction"/>
    </xs:restriction>
  </xs:simpleType>

  <xs:complexType name="Reference">
```

```
    <xs:annotation>
      <xs:documentation> Reference: External Content. The payload is the uri, and the
'type'
        describes the nature of the data under that uri. The Reference can also be named.
      </xs:documentation>
    </xs:annotation>
    <xs:sequence/>
    <xs:attribute name="uri" type="xs:anyURI" use="required"/>
    <xs:attribute name="type" type="xs:string"/>
    <xs:attribute name="mimetype" type="xs:string"/>
    <xs:attribute name="meaning" type="xs:anyURI"/>
  </xs:complexType>

</xs:schema>
```